\documentclass[aps,prl,twocolumn]{revtex4}
\usepackage{graphicx}
\usepackage{amsmath}
\usepackage{amssymb}
\usepackage{braket}
\usepackage{hyperref}
\usepackage{subfigure}
\usepackage[usenames,dvipsnames]{color}

\usepackage{graphicx}
\usepackage{subfigure}
\usepackage{slashed}
\usepackage{relsize}

\newcommand{\ct}{\cite}

\newcommand{\bi}{\bibitem}
\newcommand{\be}{\begin{equation}}
\newcommand{\ee}{\end{equation}}
\newcommand{\ba}{\begin{eqnarray}}
\newcommand{\ea}{\end{eqnarray}}

\newcommand{\non}{\nonumber}

\begin{document}

\title{Exact results  in  Floquet coin toss for  driven integrable models}
\author{Utso Bhattacharya, Somnath Maity, Uddipan Banik and Amit Dutta \\
Department of Physics, Indian Institute of Technology, Kanpur-208016, India}
\begin{abstract}
We study an integrable Hamiltonian   {reducible to free fermions}  which is  subjected to an imperfect periodic driving  with the amplitude of driving (or kicking)  randomly chosen from a binary distribution like a coin-toss
problem. The randomness present  in the driving protocol  destabilises  the periodic steady state, reached in the limit of perfectly periodic driving,  leading
to a  monotonic rise of the stroboscopic residual energy  with the number of periods ($N$). We establish  that a minimal deviation from the perfectly periodic driving would always result in
a {\it bounded} heating up of the system  with $N$ to an asymptotic finite value.   Remarkably, exploiting the completely uncorrelated nature of the randomness and the knowledge of the stroboscopic Floquet
operator in the perfectly  periodic situation, we provide an exact analytical formalism to derive the disorder averaged  expectation value of   the residual energy
 through a  {\it disorder operator}.  This formalism   not only leads to an immense numerical simplification, but also enables us to derive an exact analytical
 form for the residual energy in the asymptotic limit which is universal, i.e, independent of the  bias of coin-toss  and the protocol chosen. {Furthermore, this
 formalism  clearly establishes the nature of the  monotonic growth of the residual energy at intermediate  $N$ while clearly  revealing the possible non-universal behaviour of the same.}
\end{abstract}
\maketitle


 The study of  non-equilibrium dynamics of closed quantum systems is an exciting as well as a challenging area of recent research both from experimental \ct{bloch08,lewenstein12,jotzu14,greiner02,kinoshita06,gring12,trotzky12,cheneau12,fausti11,rechtsman13, schreiber15} and theoretical perspectives \ct{calabrese06,rigol08,oka09,kitagawa10,
lindner11,bermudez09,patel13,thakurathi13,pal10,nandkishore15,heyl13,budich15,sharma16}. One of the prominent areas in this regard is periodically driven
closed quantum systems  (for review see, \ct{dziarmaga10,polkovnikov11,dutta15,eisert15,alessio16,stat,bukov16}) which  have an illustrious history dating back to the analysis of the famous Kapitza pendulum \ct{kapitza51} and the kicked-rotor model \ct{casati79}. The recent interest in periodically driven systems are many  fold: e.g.,  Floquet engineering of materials in their non-trivial
phases such as the  Floquet graphene \ct{oka09,kitagawa10} and topological insulators \ct{lindner11} (see also \ct{cayssol13}), dynamical generation of edge Majorana \ct{thakurathi13} and non-equilibrium phase transitions
\ct{bastidas11}  like
recently proposed time crystals \ct{else16,khemani16}. These studies have received a tremendous boost following experimental studies on light-induced non-equilibrium  superconducting and topological systems \ct{fausti11,rechtsman13} and possibility of realising time crystals \ct{zhang17,choi17}. The other relevant  question deals with  fundamental statistical aspects, namely the  thermalisation of a closed
quantum system under a periodic driving \ct{russomanno12} and the possibility of the many-body localisation \ct{ponte15}.

Periodically driven closed quantum systems, from a statistical viewpoint,  are being extensively studied in the context of defect and residual energy generation  \ct{mukherjee08,russomanno12}, dynamical freezing \ct{das10}, many-body energy localization \ct{alessio13} dynamical localisation\ct{nag14,agarwala16}, dynamical fidelity \ct{sharma14}, work-statistics \ct{anirban_dutta14,russomanno15},   and as well as
in the context of entanglement entropy \ct{russomanno16} and associated dynamical phase transitions \ct{sen16}.
For  periodically driven closed integrable systems,   {reducible to free fermions},  it is usually believed that the system reaches a periodic steady state in the asymptotic limit of driving
and hence stops absorbing energy \ct{russomanno12,gritsev17}: the resulting steady state can be viewed as a periodic Gibbs ensemble
with an extensive number of conserved quantities \ct{lazarides14}.

However, for a non-integrable model \ct{alessio14}
or for an aperoidic driving of an integrable model, the system is expected  to absorb energy indefinitely; also there exists a  possibility of a geometrical generalised Gibbs ensemble in
some special situations \ct{nandy17}. However, for a driven non-integrable system a MBL state may also arise \ct{ponte15} and also a MBL state may get delocalised under a periodic driving \ct{lazarides15}.
{{While for a periodically driven non-integrable system
it is challenging to prevent the system from heating up,
the noninteracting case is fundamentally stable and does not suffer
from a ``heat death problem".   {Considering an integrable model that is  reducible to free fermions, we show below that even in this
simplest situation, the slightest deviation  from the perfect periodicity, which is experimentally inevitable, would always result in
heating up of the system  {with the stroboscopic time to a finite bounded asymptotic value. This bounded nature is in sharp contrast to the situation mentioned in Ref. \cite{ott84,guarneri84}, where it has been proved, based on an approach of the spectral analysis of a non-random operator \ct{guarneri84}, that the expectation value of the kinetic energy operator of a noisy $\delta$-perturbed quantum rotator
is unbounded in time.
}
It should be noted  at the outset that the fate of other types of integrable systems (e.g., Bethe-integrable ones) even under  a perfectly periodic drives is still an open issue \ct{bukov}.

 {Furthermore, this problem of aperiodic driving  from a broad scenario raises a plethora of pertinent questions: 
 for example, what 
would be the fate of an emergent topological phase under such a temporal noise  \ct{leon13}? Can the aperiodicity be visualised as an outcome of coupling the system to a bath \ct{iwahori16}?
What would happen to the stroboscopic entanglement entropy in the asymptotic limit \ct{russomanno16}? Can the deviation from periodicity in driving also induce localization-delocalization transitions in MBL systems
\ct{lazarides15}? Can the problem
be connected to a quantum random walk problem and corresponding search algorithms \ct{kempe03,oka03}? Although our attention here is limited to the problem of heating up only, we believe that the framework we develop would definitely provide the key foundation to address most of the questions that have been mentioned above.}
 }

In this paper, we consider a closed integrable quantum system undergoing an {\it imperfect} periodic dynamics. The imperfection or disorder manifests itself in the amplitude
of the periodic drive which assumes binary values chosen from a binomial distribution resembling a series of biased {\it coin toss} events. The combination of a periodic
driving with such a disordered amplitude results in the so-called ``Floquet coin toss problem". The aim of the paper is to  provide an exact analytic framework to  explore the statistical
properties of such a non-equilibrium system observed at $N$-th stroboscopic interval determined by the inverse of the frequency ($\omega$) of the perfectly periodic  drive. 
Our study indeed confirms that the system, even though integrable,  never  reaches a periodic steady state and rather keeps on absorbing
energy till the asymptotic limit.

We establish the above claim  considering the  one dimensional transverse Ising Model, described by the Hamiltonian,   
\begin{equation}\label{eq:1}
H=-\sum_{n=1}^{L}\tau_n^{x}\tau_{n+1}^{x}-h\sum_{n=1}^{L}\tau_{n}^{z}.
\end{equation}
where $h$ is the transverse field and $ \tau_{n}^{i}$ $\{i=x,y,z\} $ are the Pauli spin matrices at $n^{th}$ site. 
 The Hamiltonian can be decoupled in to $2\times 2$ problems for each Fourier mode via a  Jordan-Wigner  mapping,  such that  $H= \sum_{k}^{} {H_k}$ with $ H_k=(h-\mathrm{cos}k)\sigma_z+\mathrm{sin}k\sigma_x $, where $\sigma$'s are again Pauli matrices.  
 We here use
 the anti-periodic boundary condition for even $L$
 so that
   $k=\frac{2m\pi}{L}$ with $m=-\frac{L-1}{2}, ..., -\frac{1}{2}, \frac{1}{2}, ..., \frac{L-1}{2}$.

\begin{figure*}[]
\centering
\subfigure[]{%
\includegraphics[width=.45\textwidth,height=6cm]{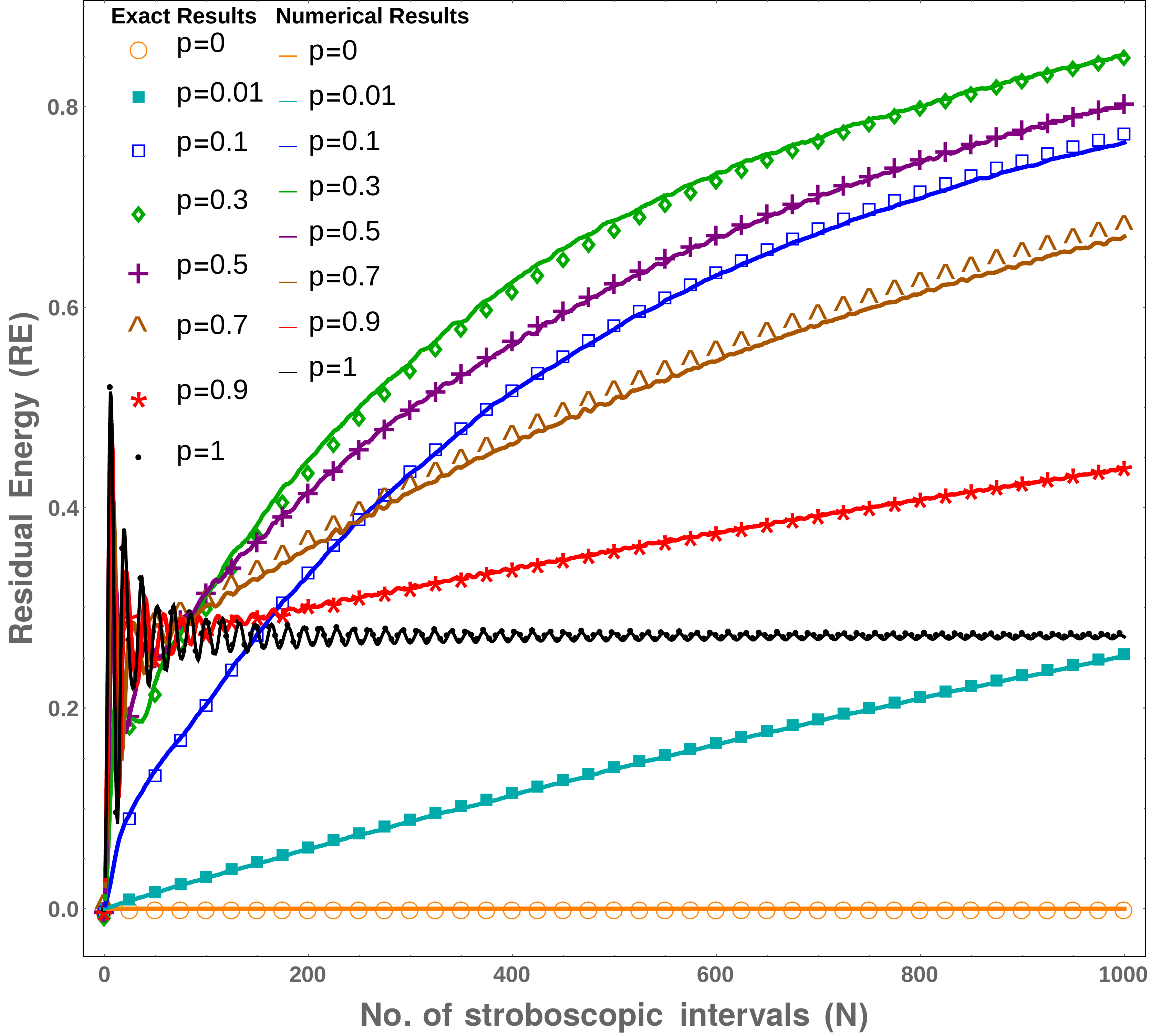}
\label{fig_delta}}
\hfill
\quad
\subfigure[]{%
\includegraphics[width=.45\textwidth,height=6cm]{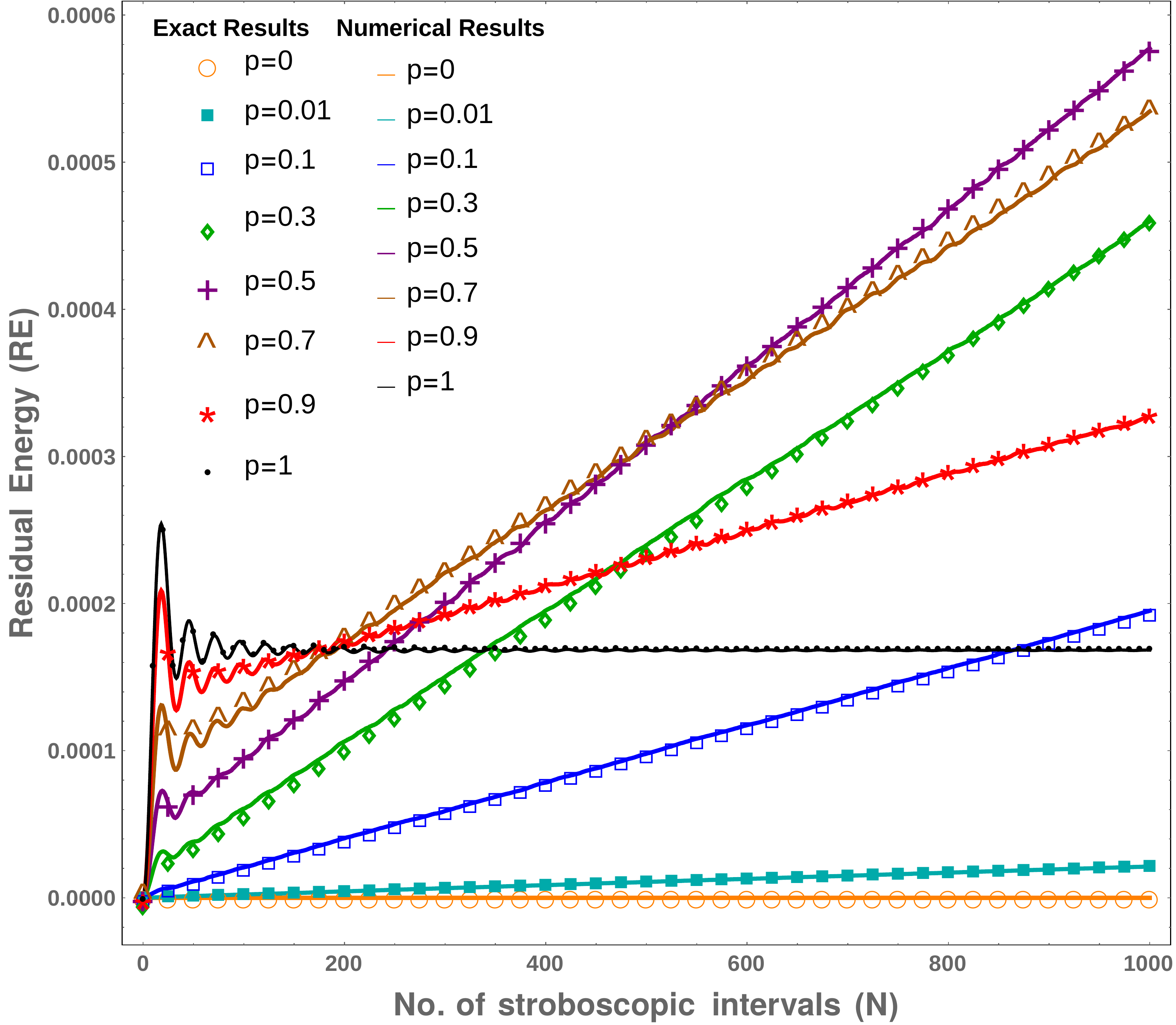}
\label{fig_sine}}
\caption{ (Color online) {(a) The RE plotted as a function of stroboscopic intervals ($N$) for a randomly kicked transverse Ising chain for $\alpha=\pi/16$. The solid lines correspond to numerically obtained results for different values of $p$, while  different symbols represent corresponding  exact analytical results.   (b)  The same  for the  random sinusoidal driving for $\alpha=1$. In both the cases, driving frequency
$\omega$ is chosen to be 100 and the number of configurations used in numerics $N_c = 1000$. For the fully periodic situation ($p=1$), the system synchronises with the external driving and stops absorbing energy. On the contrary for any non-zero
value of $p \neq 1$, the periodic steady state gets destabilised and the system keeps on absorbing heat (see text).
{We note that the nature of the  growth of the RE as a function of $N$ is not universal and rather depends on the bias $p$ and also
the protocol (see SM).}}}
\label{fig_resvn}
\end{figure*}

In our present work we study the effect of aperiodic temporal variation of the external transverse field $h$, considering two different types of driving protocols, namely, the  delta kicks and  the sinusoidal driving incorporating a binary disorder in the amplitude of driving. In short, we have $h(t)= 1 + f(t)$ 
with
\begin{align}\label{eq:3}
f(t) & = \sum_{n=1}^{N} g_n \left[ \alpha \delta (t-nT) \right] \quad  \text{ \hspace{0.05cm} for the delta kick},   &&\\ \label{eq:4}
& = \sum_{n=1}^{N} g_n  \alpha \sin \left(\frac{2\pi t}{T} \right) \quad\text{for the sinusoidal drive}, \quad
\end{align} 
where $\alpha$ is the amplitude of driving and $T$ is the time period. The random variable $g_n$ (  {where $n$ refers to the $n$-th stroboscopic period}) takes the value either $1$ with probability $p$ or $0$ with probability $(1-p)$ chosen  from a  Binomial distribution. Evidently $g_n=0$, corresponds to free evolution in the $n^{th}$ time period within the time interval  $(n-1)T$ to $nT$ while 
$g_n=1$ corresponds to the periodic perturbation in the form of a kick or sinusoidal driving. 

To illuminate the underlying Floquet theory,  let us  first consider the case of fully periodic situation $(p=1)$ choosing the initial  state  $|\psi_k(0)\rangle$  as the ground-state of the free Hamiltonian in our case; 
for each $k$ mode, we then have  $H_k(t+T)=H_k(t)$. Using the Floquet formalism, one can define a Floquet evolution operator $\boldmath{\cal{F}}_k(T) = \cal{T}$$ \exp \left( -i \int_{0}^{T}H_k(t) dt \right)$, where $\cal{T}$ denotes the time ordering operator and  the solutions of the Schr\"{o}dinger equation for a time periodic Hamiltonian can be written as, $\lvert \psi_k^{(j)}(t) \rangle = \exp(-i \epsilon_k^{(j)} T) \lvert \phi_k^{(j)}(t) \rangle$. The states $\lvert \phi_k^{(j)}(t) \rangle$, the so called Floquet modes satisfying the
condition $\lvert \phi_k^{(j)}(t+T) \rangle =\lvert \phi_k^{(j)}(t) \rangle $ and the real quantities $\epsilon_k^{(j)}$ are known as Floquet quasi-energies. In the case of the $\delta$-function kick, $\boldmath{\cal{F}}_k$ can be exactly written, in the form of,
\begin{equation}
\bold{\cal{F}}_k(T) = \exp(-i \alpha \sigma_z)\exp(-i H^0_{k}T)
\end{equation}
consisting of two pieces; the first one corresponds to  the $\delta$-kick at time $t=T$ while the second part represents the free evolution generated by the time independent Hamiltonian $H^0_k=(1-\cos k)\sigma_z + (\sin k )\sigma_x$ from time $0$ to $T$. However for the sinusoidal driving  $\bold{\cal{F}}_k(T)$ can be numerically diagonalized to obtain the quasi-energies $\epsilon_k^{\pm}$ and the corresponding Floquet modes $\rvert \phi_k^{\pm}\rangle$. Focussing on the mode $k$, after a  time $t=NT$ we have  $\lvert \psi_k(NT) \rangle = \sum_{j=\pm} r_k^{(j)} e^{-i \epsilon_k^{(j)} NT} \lvert \phi_k^{(j)}(t) \rangle$, where $r_k^{\pm}= \langle \phi_k^{\pm}\rvert \psi_k(0) \rangle $ and hence the residual energy $\varepsilon_{res}(NT)= \frac{1}{L} \sum_{k} \left( e_k(NT) - e^g_k(0) \right)$ with  $e_k(NT)=\langle \psi_k(NT) \rvert H_k^0\lvert \psi_k(NT) \rangle $ and $e^g_k(0)=\langle \psi_k(0) \rvert H_k^0 \lvert \psi_k(0) \rangle $.  In the thermodynamic limit of $L\rightarrow \infty$, we have

\begin{eqnarray}\label{eq:6}
	&~&\varepsilon_{res}(NT)  =   \frac{1}{2\pi}\int dk\biggl[\sum_{\alpha = \pm} | r_k^{\alpha} |^2 \langle \phi_k^{\alpha} \rvert H_k^0 \lvert \phi_k^{\alpha} \rangle \nonumber \\
	& + &\sum_{\substack{ \alpha,\beta = \pm \\\alpha \neq \beta}} \left(r_k^{\alpha\ast}  r_k^{\beta} e^{i\left( \epsilon_k^\alpha - \epsilon_k^\beta \right) NT } \langle \phi_k^{\alpha} \rvert H_k^0 \lvert \phi_k^{\beta} \rangle\right) -e_k^g(0) \biggr] \nonumber \\
	\label{eq_eres}
\end{eqnarray}
In the limit $N \rightarrow \infty$, due to the Riemann-Lebesgue lemma the   rapidly oscillating  off-diagonal terms in Eq. \eqref{eq_eres} drop off  upon integration over all $k$ modes leading to a steady state  expression for $\varepsilon_{res}$ \ct{russomanno12}.

Deviating from the completely periodic case and considering the situation $0 < p < 1$, so that we have a probability $(1-p)$ of missing a kick (or a cycle of sinusoidal drive is absent) in every complete period, one can now write the corresponding evolved state after $N$ complete periods as
\begin{equation}
\lvert \psi_k(NT)\rangle = \bold{U}_k(g_N) \bold{U}_k(g_{N-1}). . .  . . . .  \bold{U}_k(g_2)  \bold{U}_k(g_1) \lvert \psi_k(0) \rangle
\end{equation}
with the generic evolution operator  given by,
\begin{equation}
\bold{U}_k(g_n)=\begin{cases}
\boldmath{\cal{F}}_k(T), & \text{if $g_n = 1$}.\\
U^{0}_k(T), & \text{if $g_n = 0$}.
\end{cases}
\label{eq_protocol}
\end{equation}
where $\boldmath{\cal{F}}_k(T)$ is the usual Floquet operator and $U^0_{k}(T)= \exp(-i H^0_{k} T)$ is the time evolution operator for the free Hamiltonian $H^0_{k}$.  One then readily finds:
\begin{flalign} 
e_k(NT) & =  \langle \psi_k(0) \rvert \bold{U}_k^{\dagger}(g_1) \bold{U}_k^{\dagger}(g_2) ....... \bold{U}_k^{\dagger}(g_{N-1}) \bold{U}_k^{\dagger}(g_N) &&\nonumber \\
&\times H^0_{k} \bold{U}_k(g_N) \bold{U}_k(g_{N-1})........\bold{U}_k(g_2) \bold{U}_k(g_1) \lvert \psi_k(0) \rangle
\label{eq_rs1}
\end{flalign}
The numerical calculation in Eq.~\eqref{eq_rs1} thus involves   {two steps:  (i) the multiplication of $(2N+1)$   matrices corresponding to $N$ complete periods
for a given disorder configuration as shown in Eq.~\eqref{eq_rs1}. (ii) Configuration averaging over the disorder for a fixed  value of $N$}. Let us refer  to the Fig.~\ref{fig_resvn}, where numerically obtained residual energy (RE) is plotted as a function of number of complete periods $N$ choosing a high frequency
limit ($\omega=2\pi/T >4$)  in which the fully periodic situation leads to a periodic steady state both for the $\delta$-kicks and sinusoidal variation and the system
synchronises with the external driving. 
 Observing the monotonic rise of RE with $N$ for a given $p\neq 0,1$, we conclude that for any non-zero value of $p$, the periodic steady state gets destabilised and system keeps absorbing energy. 
 We further note, that for a given $N$, the behavior of the RE cannot monotonically rise with $p$ since the RE gets constrained by the fully periodic situation around $p=1$ and the no rise situation around $p=0$.
 The value of $p$ for which the rate of rise of RE with $N$ will be  maximum depends on both $\alpha$ and $\omega$ of the imperfect drive (see supplementary material (SM)) \ 
 {Furthermore, a slightest inclusion of aperiodicity in the drive leads to the difference in the RE to rise linearly with $N$ (up to an appropriate value of $N$) both in the limit $p \to 0$ from the undriven situation or $(1-p) \to 0$ from the perfectly periodic driven situation,  as has been elaborated explicitly in the SM. }


Having provided the numerical results, we shall proceed to set up the corresponding  analytical framework within the space spanned by  the complete set of 
Floquet basis $ \{ \lvert j_k^{\pm} \rangle\} $ states.
Introducing $2(N+1)$  identity operators in terms of  the Floquet basis  states $\sum_{j^{(m)}} \lvert j^{(m)}_k \rangle \langle j^{(m)}_k \rvert =  \hat {1}$ in Eq.~\eqref{eq_rs1}, where $j^{(m)}_k$  can take two possible values corresponding to two quasi-states of the $2 \times 2$ Floquet Hamiltonian $\boldmath{\cal{F}}_k(T)$ for each mode $k$ and performing the average over disordered configurations and finally upon reorganisation, we find,  
\begin{widetext}
\begin{eqnarray}
 \langle e_k(NT) \rangle =  \slashed\sum \langle \psi_k(0) \rvert j^{(0)}_{k} \rangle \langle j^{(N)}_{k} \rvert H^0_{k} \lvert i^{(N)}_{k} \rangle \langle i^{(0)}_{k} \lvert \psi_k (0) \rangle 
 \left[ \prod_{m=1}^{N} \left( \sum_{g_m = 1, 0} P(g_m) \langle j^{(m-1)}_{k} \rvert \bold{U}_k^{\dagger}(g_m) \lvert j^{(m)}_{k} \rangle \langle i^{(m)}_{k} \rvert \bold{U}_k(g_m) \lvert i^{(m-1)}_{k} \rangle \right) \right] \nonumber\\
 \end{eqnarray}
 \end{widetext}
where $\slashed \sum \equiv \sum_{\substack{ j^{(0)},j^{(1)},....,j^{(N)} \\i^{(0)}, i^{(1)}, ....,i^{(N)}}}$. 
The uncorrelated nature of   $g_m$s enables us to perform the  configuration average by  separately  averaging over  for each  $g_m$. Recalling Eq.~\eqref{eq_protocol},  and the fact that $\boldmath{\cal{F}}_k(T) |j^\pm_k\rangle=\exp(-i\epsilon^\pm_kT) |j^\pm_k\rangle$ and $P(g_m) $ is the probability of being perfectly driven [$P(g_m =1) = p$] and free evolution [$P(g_m=0)=(1-p)$], respectively, leads us to:
 \begin{widetext}
 \begin{eqnarray}
&&\langle e_k(NT) \rangle=  \slashed \sum  \langle \psi_k(0) \rvert j^{(0)}_{k} \rangle \langle j^{(N)}_{k} \rvert H^0_{k} \lvert i^{(N)}_k \rangle \langle i^{(0)}_k \lvert \psi_k(0) \rangle \nonumber\\
& \times& \left[ \prod_{m=1}^{N}  \left( p e^{iT\left(\epsilon_k^{j_{m-1}}-\epsilon_k^{i_{m-1}}\right)} \delta_{j^{(m-1)}_k,j^{(m)}_k} \delta_{i^{(m)}_k,i^{(m-1)}_k} + (1-p)   \langle j^{(m-1)}_k \rvert U^{0\dagger}_k \lvert j^{(m)}_k \rangle \langle i^{(m)}_k \rvert U^0_k \lvert i^{(m-1)}_k \rangle  \right) \right]\nonumber \\
 &=& \sum_{\substack{ j^{(0)},j^{(N)} \\i^{(0)},i^{(N)}}} \langle \psi_k(0) \rvert j^{(0)}_k \rangle \langle j^{(N)}_k \rvert H^0_k \lvert i^{(N)}_k \rangle \langle i^{(0)}_k \lvert \psi_k(0) \rangle \left(  \sum_{\substack{ j^{(1)},j^{(2)}.....,j^{(N-1)} \\i^{(1)},i^{(2)},.....,i^{(N-1)}}}  \prod_{m=1}^{N} D^{j^{(m-1)},j^{(m)},i^{(m-1)},i^{(m)}}_k \right)\nonumber \\
&=& \sum_{\substack{ j^{(0)},j^{(N)} \\i^{(0)},i^{(N)}}} \langle \psi_k(0) \rvert j^{(0)}_k \rangle  \langle j^{(N)}_k \rvert H^{(0)}_k \lvert i^{(N)} \rangle \langle i^{(0)}_k \lvert \psi_k(0) \rangle \Big[ \bold{D}_k^N \Big]_{j^{(0)},j^{(N)},i^{(0)},i^{(N)}}
\label{eq_final_eres}
\end{eqnarray} 
\end{widetext}
where the matrix element of ($4\times4$) matrix $\bold{D}_k$ is given by,
\begin{widetext}
\begin{align}
\bold{D}^{j^{(m-1)},j^{(m)},i^{(m-1)},i^{(m)}}_k \equiv \left(p e^{iT\left(\epsilon^{j_{m-1}}_k-\epsilon_k^{i_{m-1}}\right)} \delta_{j^{m-1}_k,j^{m}_k} \delta_{i^{m}_k,i^{m-1}_k} + (1-p)   \langle j^{(m-1)}_k \rvert \bold{U}_k^{0\dagger} \lvert j^{(m)}_k \rangle \langle i^{(m)}_k \rvert \bold{U}^0_k \lvert i^{(m-1)}_k \rangle \right)
\label{eq_dmat}
\end{align} 
\end{widetext}
It should be noted that in Eq.~\eqref{eq_final_eres}, the $N$ in $\bold{D}_k^N$ is not a label but the matrix $\bold{D}_k$ that we have defined in Eq.~\eqref{eq_dmat}, raised to the power $N$. The above exercise naturally leads to the emergence  of  $4 \times 4$ disorder matrix $\bold{D}$ for a imperfectly driven $2 \times 2$ system. Given
the amplitude, frequency, dimensionality, the form  of  ${\mathcal F}_k(T)$ in every stroboscopic intervals and the knowledge of disorder encoded in the probability $p$
of driving, every element of $\bold{D}$-matrix can be exactly calculated as shown {in the SM
  {where   we probe the analytical structure of the disorder matrix  emphasising both on  the limit $N \to \infty$ and intermediate $N$.} The intriguing feature is   that   one of the eigenvalues of the (non-unitary) disorder matrix becomes  unity   while the other real  eigenvalue and also the modulus of
complex eigenvalues are less than unity and thus  become vanishingly small in the diagoanlised form of ${\mathbf {D}}^N$ in the limit $N\to \infty$. This remarkable  property of the $\mathbf D$-matrix immediately renders an exact analytical form of the residual energy in the asymptotic limit}:

\begin{equation}
\lim_{N\to\infty}{\langle\varepsilon_{res}(NT)\rangle}=\frac{1}{\pi}\left[e^g_{k=\pi}(0) -\int_0^\pi{e^g_k(0) dk}\right].
\label{eq_supres3}
\end{equation}

{The asymptotic value obtained in Eq.~\eqref{eq_supres3} is  clearly  independent of the driving strength, frequency and the protocol implemented, and hence   is evidently {\it universal}. It is also finite and bounded in contrast to the kicked rotor case in Ref. \cite{guarneri84}. Further, the RE as obtained from the exact form given in  Eq. \eqref{eq_final_eres} confirms the numerically obtained values presented in Fig. \ref{fig_resvn} for all values of $N$.
}

{For small values of $N$, when the transients have not yet decayed, the slope of the residual energy curves for the two different protocols (as shown in Fig. \ref{fig_resvn}) are different. This difference, however disappears in the long time limit when  the initial transients die off.  Hence, the curves for residual energy for both the protocols exhibit a universal behaviour at large but intermediate $N$ as  depicted in Fig. \ref{fig_comparison}.  This  coarse-grained (in time) view  of $\varepsilon_{res}$ establishes  the similar behaviour for two protocols in such large time scale; this can interestingly be understood by noting that at large $N$, the phase of the complex eigenvalues, $\exp[\pm i N\phi(k,\alpha,\omega)]$, of the disorder operator, which are responsible for the initial interferences, oscillate rapidly and becomes vanishingly
small for large intermediate  $N$ as illustrated in the SM. However, the other real eigenvalue $r=r(k,\alpha, \omega)$, with modulus less than unity survives for large intermediate  $N$ and goes to zero only when $N\to\infty$. Thus, out of the four eigenvalues of the disorder matrix only two eigenvalues, $1$ and $r$ (see SM) raised to the power $N$, are non-vanishing and compete against each other. While the unit eigenvalue forces the system to achieve an universal value at $N\to\infty$, $r^N$ at large intermediate times is solely responsible for the  monotonic rise of the residual energies for any protocol chosen. Of course, $r^N \to 0$ eventually at $N\to\infty$. Thus, the monotonic growth of $\varepsilon_{res}$ with increasing $N$, is ensured by the gradual decay of the (positive) $r^N$. We note that although at finite $N$ the curves at long time scales are similar, their slopes as they tend towards the asymptotic limit  are different for the two protocols. This can be ascribed to the fact that the eigenvalue $r=r(k,\alpha,\omega)$ is not only dependent on the driving strength and frequency but its functional form is dictated by the protocol chosen and hence, its value is non-universal.}
Thus, we have constructed a complete analytic framework to deal with  any ``Floquet
coin toss" problem in general.  

The numerical simulations addressing similar problems involving multiplication of $N$ unitary matrices, followed by  averaging over  a large number  ($N_c$) of disorder configurations; as  $N$ and $N_c \to \infty$, requires huge computational time while being susceptible to numerical inaccuracies. {On the contrary, the non-perturbative $\mathbf {D}$-matrix, overcomes such numerical challenges by calculating the exact analytical behavior of $\varepsilon_{res}$ at large $N$
and especially in the asymptotic limit which is otherwise extremely expensive  numerically \cite{comment} .}

\begin{figure}[]
\includegraphics[width=.45\textwidth,height=6cm]{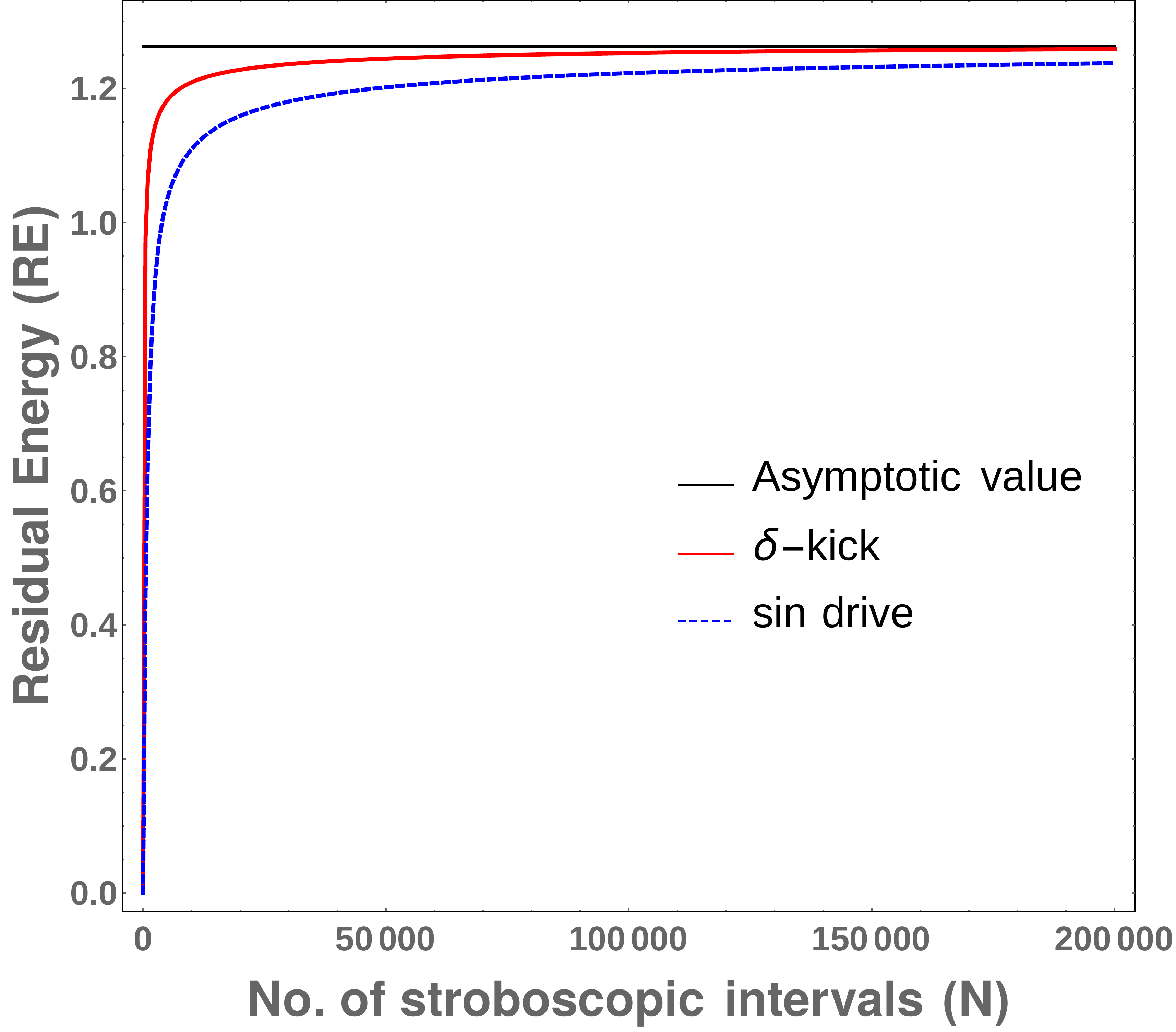}
\caption{ (Color online) 
An \textit{exact} coarse-grained view of  $\langle \varepsilon_{res}(NT) \rangle$  as a function of the stroboscopic interval $N$ as obtained from the disorder operator formalism with $L=100$, $\omega=100$, $p=0.5$ for $\delta$-kicked ($\alpha=\pi/2$ in solid red) and  for the sinusoidal variation ($\alpha= 1000$ in dotted blue) \ct{comment}. The asymptotic value is the horizontal line on top  plotted in solid black. The nature of the curve is the same for both the protocols and is universal while the slopes differ as an artefact of the real positive eigenvalue $r(k,\alpha,\omega)$ being non-universal.}
\label{fig_comparison}
\end{figure}

  {The sole aim of this paper has been to develop the disorder matrix formalism, given the complete stroboscopic information for a perfectly periodic case, to establish the bounded growth of the residual energy to a asymptotic value following an aperiodic driving  keeping free fermion reducible integrable quantum systems in mind. However, we believe it can be generalised to situations where the disorder matrix is higher dimensional. In such scenarios, the asymptotic value of the residual energy and other such operators may still be obtained from our analysis. For non-integrable situations, given uncorrelated binary distribution of disorder, if complete stroboscopic information 
  about the Floquet evolution operator for
the $p=1$ situation  
is completely known, the disorder matrix formalism is expected to hold.}      {We note that the aperiodic  kicking situation proposed  here has already  been experimentally  realised for a single rotor by  Sarkar $et ~al.$ \ct{sarkar17};  
similar experimental studies  for aperiodically driven many body systems are indeed possible}.  Given a rare possibility of analytical approach to explore a temporally disordered situation
and given the wide scope of the validity of our results, we believe that our approach is going to provide a new avenue to a plethora
of similar studies. {We conclude with a note that while spatial disorder leads to Anderson localisation, the temporal
disorder which we study here remarkably leads to  a delocalisation (in the Floquet space) with a bound; the importance of our work in this regard has already been noted in Ref. \cite{quelle17}.}
\begin{acknowledgments}
AD acknowledges SERB, DST, New Delhi for financial support. We acknowledge Diptiman Sen for critical comments.
\end{acknowledgments}

\widetext
\begin{center}
	\textbf{\large Supplementary Material on ``Exact results  in  Floquet coin toss for  driven integrable models''}\\
	\vspace{0.5cm}
	{  Utso Bhattacharya, Somnath Maity, Uddipan Banik  and Amit Dutta }\\
	\vspace{0.2cm}
	{}{\it Indian Institute of Technology Kanpur, Kanpur 208 016, India} \\
\end{center}

\setcounter{equation}{0}
\setcounter{figure}{0}
\setcounter{table}{0}
\setcounter{page}{1}
\makeatletter
\renewcommand{\theequation}{S\arabic{equation}}
\renewcommand{\thefigure}{S\arabic{figure}}
\renewcommand{\bibnumfmt}[1]{[S#1]}
\renewcommand{\@cite}[1]{[S#1]}

\section{The analytical structure of the disorder matrix and the growth of the residual energy}

In this supplementary material, we shall probe the analytical structure of the disorder matrix and thus establish the  growth
of the residual energy as a function of the stroboscopic period $N$, especially emphasising the limit $N \to \infty$ and large intermediate $N$. What is intriguing, as we shall show below, is that  one of the eigenvalues of the (non-unitary) disorder matrix is  unity while the other eigenvalues ( real and the modulus of
the two complex eigenvalues) are less than unity
and thus  become vanishingly small in the diagonalized form of $D^N(k)$ in the limit $N\to \infty$. This remarkable  property of the $D$-matrix results
in a very simplified form of the disorder matrix which immediately renders an exact analytical form of the residual energy in the asymptotic limit which is independent of the value of the bias $p$ and the protocol used.
For finite $N$ on the other hand, the growth of the residual energy shows a non-universal behaviour, i.e.,  is  bias and protocol dependent as in this limit its behaviour is dictated by both the real eigenvalues. 

\subsection{The exact analytical structure of the disorder matrix}
To present the exact analytical structure of the $D$-matrix, let us  consider the case of a 1-D Ising model subjected to an imperfect  driving protocol referring to Eq.~(11)
of the main text:

\begin{align}
{D}^{j^{(m-1)},j^{(m)},i^{(m-1)},i^{(m)}} (k) \equiv \left(p e^{iT\left(\epsilon^{j_{m-1}}_k-\epsilon_k^{i_{m-1}}\right)} \delta_{j^{m-1}_k,j^{m}_k} \delta_{i^{m-1}_k,i^{m}_k} + (1-p)   \langle j^{(m-1)}_k \rvert \bold{U}_k^{0\dagger} \lvert j^{(m)}_k \rangle \langle i^{(m)}_k \rvert \bold{U}^0_k \lvert i^{(m-1)}_k \rangle \right)
\label{eq_dmat}
\end{align}

We recall   that $\epsilon_k^j$ are the quasi-energies  and $\ket{j_k^{1,2}}$ are corresponding eigenvectors of the $2\times 2$ Floquet Hamiltonian. On the other hand,
$U_k^0(T)$ represents the time evolution operator of the undriven Ising Hamiltonian measured at stroboscopic instant $T$.
The matrix representation of $U_k^0(T)$ is as follows:
\begin{equation}
U_k^0(T) \doteq \left(
\begin{array}{cc}
u_{11}(k)-i u_{12}(k) & -i u_{21}(k) \\
-i u_{21}(k) & u_{11}(k)+i u_{12}(k) \\
\end{array}
\right)
\end{equation}
where
\begin{eqnarray}
u_{11}(k)&=&\cos \left(2 T \sin \left(\frac{k}{2}\right)\right)\\
u_{12}(k)&=&\sin \left(\frac{k}{2}\right) \sin \left(2 T \sin \left(\frac{k}{2}\right)\right)\\
u_{21}(k)&=&\cos \left(\frac{k}{2}\right) \sin \left(2 T \sin \left(\frac{k}{2}\right)\right)
\end{eqnarray}

{Due to the unitarity, the matrix $U_k^0$} in the basis of the eigenvectors $\ket{j_k^{1,2}}$ can be cast in a general form:
\begin{equation}
\bra{j_k^m}U_k^0(T)\ket{j_k^n} \doteq \left(
\begin{array}{cc}
W_{11}(k) & W_{12}(k) \\
-W_{12}^*(k) & W_{11}^*(k) \\
\end{array}
\right)
\label{eq_matrix2by2}
\end{equation}
with $|W_{11}(k)|^2+|W_{12}(k)|^2=1$.
Thus, a general form of the $4\times 4$ disorder matrix can easily be constructed for the above situation to obtain,

\begin{eqnarray}
D(k) &\doteq & \left(
\begin{array}{cccc}
p+(1-p) \left| W_{11}\right|^2 & -(1-p)W_{11}^*W_{12}^* & -(1-p)W_{11}W_{12} & (1-p) \left| W_{12}\right|^2 \\\\
(1-p)W_{11}^*W_{12} & p\exp{[i\Delta\phi_k T]}+(1-p)W_{11}^*W_{11}^* & -(1-p)W_{12}W_{12} & -(1-p)W_{11}^*W_{12} \\\\
(1-p)W_{11}W_{12}^* & -(1-p)W_{12}^*W_{12}^* & p\exp{[-i\Delta\phi_k T]}+(1-p)W_{11}W_{11} & -(1-p)W_{11}W_{12}^*\\\\
(1-p) \left| W_{12}\right|^2 & (1-p)W_{11}^*W_{12}^* & (1-p)W_{11}W_{12} & p+(1-p) \left| W_{11}\right|^2 \\
\end{array}\non
\label{eq_D_matrix}
\right) \\ 
&\doteq &\left(
\begin{array}{cccc}
r_1(k) & c_1(k) & c_1^*(k) & r_2(k) \\
c_2(k) & c_3(k) & c_4(k) & -c_2(k) \\
c_2^*(k) & c_4^*(k) & c_3^*(k) & -c_2^*(k) \\
r_2(k) & -c_1(k) & -c_1^*(k) & r_1(k)\\
\end{array}
\right)\non\\
\end{eqnarray}
where $\Delta\phi_k=\epsilon_k^{+}-\epsilon_k^{-}$, $r_i(k)$ and  $c_i(k)$ denotes the real and the complex elements of the matrix, respectively.
{Let us recall that  the disorder matrix $D(k)$ emerges  due to disorder (classical) averaging over  infinite number of configurations and hence evidently} is a non-Unitary matrix with the absolute values of all its elements being less than unity. 
{Further, in the case of perfectly periodic driving $p=1$, the off-diagonal terms of the $D$-matrix in \eqref{eq_D_matrix} vanishes rendering it in a diagonal form.}

\subsection{Eigenvalues and eigenvectors of the $D$-matrix in the limit $N \to \infty$}

{To continue our analysis further, let us first focus on the modes, $k=0$ and $\pi$;   the off-diagonal terms of the Hamiltonian $H^0_k=(1-\cos k)\sigma_z + (\sin k )\sigma_x$  vanishes for
	these modes and hence these modes do not evolve with time}.
The $2 \times 2$ matrix
in \eqref{eq_matrix2by2} becomes identity matrix for $k=0$ and a diagonal matrix with two diagonal elements $\exp(\mp 2iT)$ for $k=\pi$. Hence from \eqref{eq_D_matrix}, we immediately find

\begin{equation}
D(k=0, \pi)=\left(
\begin{array}{cccc}
1 & 0 & 0 & 0 \\
0 & p\exp{[i\Delta\phi_{k=0, \pi} T]}+(1-p)\exp{[i\Delta e_{k=0, \pi} T]} & 0 & 0 \\
0 & 0 & p\exp{[-i\Delta\phi_{k=0, \pi}+(1-p)\exp{[-i\Delta e_{k=0, \pi} T]} T]} & 0 \\
0 & 0 & 0 & 1 \\
\end{array}
\right)
\end{equation}
where $\Delta e_{k} =e_{k}^{e}(0)- e_{k}^{g}(0) $ is the energy gap of the Hamiltonian $H^0_{k}$.  For the sinusoidal drive $\Delta e_{k} = \Delta \phi_{k}$, whereas this is not true for the $\delta$-kicked situation as the integral of $\delta(t-T)$ is not zero over a complete period. Since both the matrices $D(k=0, \pi)$ is diagonal, it is straightforward to estimate $\lim_{N \to \infty} D^N$ for these modes. For the sinusoidal drive,

\begin{equation}
\lim_{N\to\infty}D^N(k=0)=\left(
\begin{array}{cccc}
1 & 0 & 0 & 0 \\
0 & 1 & 0 & 0 \\
0 & 0 & 1 & 0 \\
0 & 0 & 0 & 1 \\
\end{array}
\right);
\lim_{N\to\infty}D^N(k=\pi)=\left(
\begin{array}{cccc}
1 & 0 & 0 & 0 \\
0 & 0 & 0 & 0 \\
0 & 0 & 0 & 0 \\
0 & 0 & 0 & 1 \\
\end{array}
\right)
\label{eq_kpi}
\end{equation}
Whereas for the $\delta$-kicked situation,

\begin{equation}
\lim_{N\to\infty}D^N(k=0, \pi)=\left(
\begin{array}{cccc}
1 & 0 & 0 & 0 \\
0 & 0 & 0 & 0 \\
0 & 0 & 0 & 0 \\
0 & 0 & 0 & 1 \\
\end{array}
\right)
\label{eq_kpi_delta}
\end{equation}

For other modes ($k \neq 0,\pi$), the matrix $D(k)$  indeed has off-diagonal terms and hence to compute
the four eigenvalues $\lambda (k)$,    {one needs to  analyse the equation}:
\begin{equation}
\left[\lambda (k)-s(\lambda, k)\right]f(\lambda ,k)=0
\label{eq_eigenvalue}
\end{equation}
where $s(\lambda, k) =\left\{p+(1-p)\left( \left| W_{11}(k)\right|^2+\left| W_{12}(k)\right|^2\right)\right\}=1$ as $\left( \left| W_{11}(k)\right|^2+\left| W_{12}(k)\right|^2\right)=1$,
and $f(k,\lambda)$ is  a third degree polynomial with all real coefficients.
Thus, it is obvious that always one of the eigenvalues $\lambda (k)=\lambda_1 (k)=1$ with normalised eigenvector, $v_1=\frac{1}{\sqrt{2}}\left(
\begin{array}{c}
1 \\
0 \\
0 \\
1 \\
\end{array}
\right)$ for $k\neq 0,\pi$. 
While one eigenvalue sticks to unity,  it is straightforward to argue that other eigenvalues (one real and the other two complex conjugates of each other) will have a value (or modulus) less
than unity due to the presence of off-diagonal terms in matrix in \eqref{eq_D_matrix} and   vanish in $D^N(k\neq 0,\pi)$ when $N \to \infty$. Given the simple structure of the diagonal form of the $\lim_{N\to\infty}D^N(k\neq 0,\pi) = diag(1,0,0,0)$, it is easy to verify: 
\begin{equation}
\lim_{N\to\infty}D^N(k\neq 0,\pi)=\frac{1}{2}\left(
\begin{array}{cccc}
1 & 0 & 0 & 1 \\
0 & 0 & 0 & 0 \\
0 & 0 & 0 & 0 \\
1 & 0 & 0 & 1 \\
\label{eq_neqkpi}
\end{array}
\right)
\end{equation}


\subsection{The Residual Energy as $N\to\infty$}

In this section we show the behavior of the residual energy $\varepsilon_{res}(NT)$ in the limit of $N\to\infty$ using the matrices given in \eqref{eq_kpi} and \eqref{eq_neqkpi}; for a transverse Ising chain
one readily finds
\begin{eqnarray}
\lim_{N\to\infty}\langle e_{k=0}(NT)\rangle &=& e_{k=0}^{g}(0)=0 \\
\lim_{N\to\infty}\langle e_{k=\pi}(NT)\rangle &=& e_{k=\pi}^{g}(0)=-2 \\
\lim_{N\to\infty}\langle e_{k\neq 0,\pi}(NT)\rangle &=& \frac{1}{2} {\rm Tr}[H^0_{k\neq 0,\pi}]=0 
\end{eqnarray}

Let us recall that
$\varepsilon_{res}(NT)= \frac{1}{L} \sum_{k} \left( e_k(NT) - e_k^{g}(0) \right)$ with  $e_k(NT)=\langle \psi_k(NT) \rvert H_k^0\lvert \psi_k(NT) \rangle $ and $e_k^{g}(0)=\langle \psi_k(0) \rvert H_k^0 \lvert \psi_k(0) \rangle $; in the
thermodynamic limit ($L \to \infty$)

\begin{equation}
\varepsilon_{res}(NT)=  \frac{1}{\pi} \int_0^\pi dk \left[ e_k(NT) - e^g_k(0) \right]
\label{eq_supres1}
\end{equation}
Let us consider the trivial case $p=0$ (no periodic drive), for which 
the residual energy $\varepsilon_{res}(NT)=0$ as the system always remains  in the initial ground state (except for a trivial phase factor) and hence two terms in \eqref{eq_supres1} identically cancel each other for
all values of $N$.
On the other hand, in the perfectly periodic situation  ($p=1$), all the $k$-modes contribute and the steady state value becomes,
\begin{equation}
\lim_{N\to\infty}{\varepsilon_{res}(NT)}=\frac{1}{\pi}\int_0^\pi{dk\left[\left\{\sum_{\alpha=1,2}\left|\langle j^{\alpha}_{k}\ket{\psi^0_{k}}\right|^2\langle j^{\alpha}_{k}\left|H^0_{k}\right|j^{\alpha}_{k}\rangle\right\}-e^g_k(0)\right]}
\label{eq_supres2}
\end{equation}
It should be noted that the steady state value is attained in the asymptotic limit  following a partial  cancellation of  the initial ground state energy . 

Let us  now immediately contrast this scenario with the case when $p\neq 1,0$ when the disordered average residual energy
\begin{equation}
\lim_{N\to\infty}{\langle\varepsilon_{res}(NT)\rangle}=\frac{1}{\pi}\lim_{N\to\infty}\langle e_{k=\pi}(N)\rangle-\frac{1}{\pi}\int_0^\pi{e^g_k(0) dk}=\frac{1}{\pi}\left[e^g_{k=\pi}(0) -\int_0^\pi{e^g_k(0) dk}\right]
\label{eq_supres3}
\end{equation}
and observe that unlike in the steady state scenario in \eqref{eq_supres2}, as the system gets heated up, only the $k = \pi$ mode contributes in \eqref{eq_supres3} to the cancellation  to finally
yield a bias $p$ as well as protocol independent value.
Since, only one $k$-mode, the $k=\pi$ mode contributes, the asymptotic value of residual energy for $p\neq 1,0$ is evidently much greater than the steady state value, clearly asserting that the system has heated up to a finite asymptotic value.
Equation \eqref{eq_supres3} yields  the maximum energy that the system (after being heated up) can attain in the asymptotic limit 
which is independent of $p$ as well as the protocol.

However, one remaining question is that how does the system reach this asymptotic value  for $p \neq 0,1$ as $N$ increases; this entirely depends on how  the eigenvalues
of the $D$-matrix (other than the one that sticks to unity) decay in diagonal form of $D^N(k)$ as $N$ increases, and hence, as we illustrate below, on the protocol and the value of $p$.  This is precisely the reason we observe different initial growth of $ \langle \varepsilon_{res} (NT) \rangle$ in Fig. ~1a and 1b of the main text for different protocols and for different values of $p$ for a given protocol.

\subsection{The Residual Energy at Large Intermediate $N$}

In this subsection,  we shall show how the growth of the RE depends on the non-universal features in the disorder operator. The D-matrix in Eq.~\eqref{eq_D_matrix}, as has already been mentioned can be diagonalised to obtain four eigenvalues, out of which two eigenvalues $1$ and $r=r(k,\alpha,\omega)$ are real and positive, while the rest two $\lambda_c=|\lambda_c(k,\alpha,\omega)|e^{\pm i\phi(k,\alpha,\omega)}$ are complex conjugates of each other with modulus less than unity. The three roots other than one are obtained by solving $f(k,\lambda)=0$, where
$f(k,\lambda)$ is  a third degree polynomial (see Eqs.~\eqref{eq_D_matrix} and \eqref{eq_eigenvalue}) with all real coefficients and is of the form $f(k,\lambda)=\lambda^3+a_2\lambda^2+a_1\lambda+a_0$:

where 
\begin{eqnarray}
a_2&=&-\big\{2\operatorname{Re}\left[c_3\right]+(r_1-r_2)\big\}\\
a_1&=&-\big\{4\operatorname{Re}\left[c_1c_2\right]+(|c_4|^2-|c_3|^2)-2\operatorname{Re}\left[c_3\right](r_1-r_2)\big\}\\
a_0&=&\big\{4\operatorname{Re}\left[c_1c_2c^*_3\right]-4\operatorname{Re}\left[c_1c^*_2c_4\right]+(|c_4|^2-|c_3|^2)(r_1-r_2)\big\}
\end{eqnarray}

The real root $r=r(k,\alpha,\omega)$ is,

\begin{equation}
r=-\frac{1}{3}a_2+(S^++S^-)
\end{equation}

and the two self-conjugate complex roots are,
\begin{equation}
\lambda_c^\pm=-\frac{1}{3}a_2-\frac{1}{2}(S^++S^-)+\pm i\frac{\sqrt{3}}{2}(S^+-S^-)
\end{equation}

where, $S^\pm=(R\pm D)^{\frac{1}{3}}$, $ D=Q^3+R^3$, $R=\frac{9a_2a_1-27a_0-2a_2^3}{54}$, $Q=\frac{3a_1-a_2^2}{9},  $ and $\left(r+\lambda_c^++\lambda_c^-\right)=a_2$ shows that $r$ is real.

The eigenvectors of $D(k)$ as a function of the roots $\lambda$ are given as,
\begin{equation}
\left(
\begin{array}{c}
x_1(k,\lambda) \\
x_2(k,\lambda) \\
x_3(k,\lambda) \\
x_4(k,\lambda) \\
\end{array}
\right)
\end{equation}

where,

\begin{eqnarray}
x_2(k,\lambda) &=& x_1(k,\lambda)\frac{(r_1+r_2-\lambda)\big\{c^*_2c_4+c_2\left(\lambda-c^*_3\right)\big\}}{d(k,\lambda)}\\
x_3(k,\lambda) &=& x_1(k,\lambda)\frac{(r_1+r_2-\lambda)\big\{c_2c^*_4+c^*_2\left(\lambda-c_3\right)\big\}}{d(k,\lambda)}\\
x_4(k,\lambda) &=& x_1(k,\lambda)\frac{c_1\left(c_2c^*_3-c^*_2c_4-c_2\lambda\right)-\left(r_1-\lambda\right)\left(\left(|c_3|^2-|c_4|^2\right)-2\operatorname{Re}\left[c_3\right]\lambda+\lambda^2\right)-c^*_1\big\{c_2c^*_4 +c^*_2\left(\lambda-c_3\right)\big\}}{d(k,\lambda)}\non\\
\non\\
\end{eqnarray}
and 
\begin{equation}
d(k,\lambda) = c_1\left(c_2c^*_3-c^*_2c_4-c_2\lambda\right)+r_2\left(\left(|c_3|^2-|c_4|^2\right)-2\operatorname{Re}\left[c_3\right]\lambda+\lambda^2\right)-c^*_1\big\{c_2c^*_4 +c^*_2\left(\lambda-c_3\right)\big\}
\end{equation}

Imposing the orthonormality condition over the eigenvectors of $D(k)$ one can write down the diagonalising matrix $S(k,\alpha,\omega)$ as: 
\begin{equation}
S=\left(
\begin{array}{cccc}
1 & -x^*_1(\lambda^+_c) & -x^(\lambda^+_c) & -x_1(r) \\
0 & x_2(\lambda^+_c) & x_3(\lambda^+_c) & -x^*_2(r) \\
0 & x_3(\lambda^+_c) & x^*_2(\lambda^+_c) & -x_2(r) \\
1 & x^*_1(\lambda^+_c) & x_1(\lambda^+_c) & x_1(r) \\
\end{array}
\right)
\end{equation}

\begin{figure*}[]
	\centering
	\subfigure[]{%
		\includegraphics[width=.45\textwidth,height=6cm]{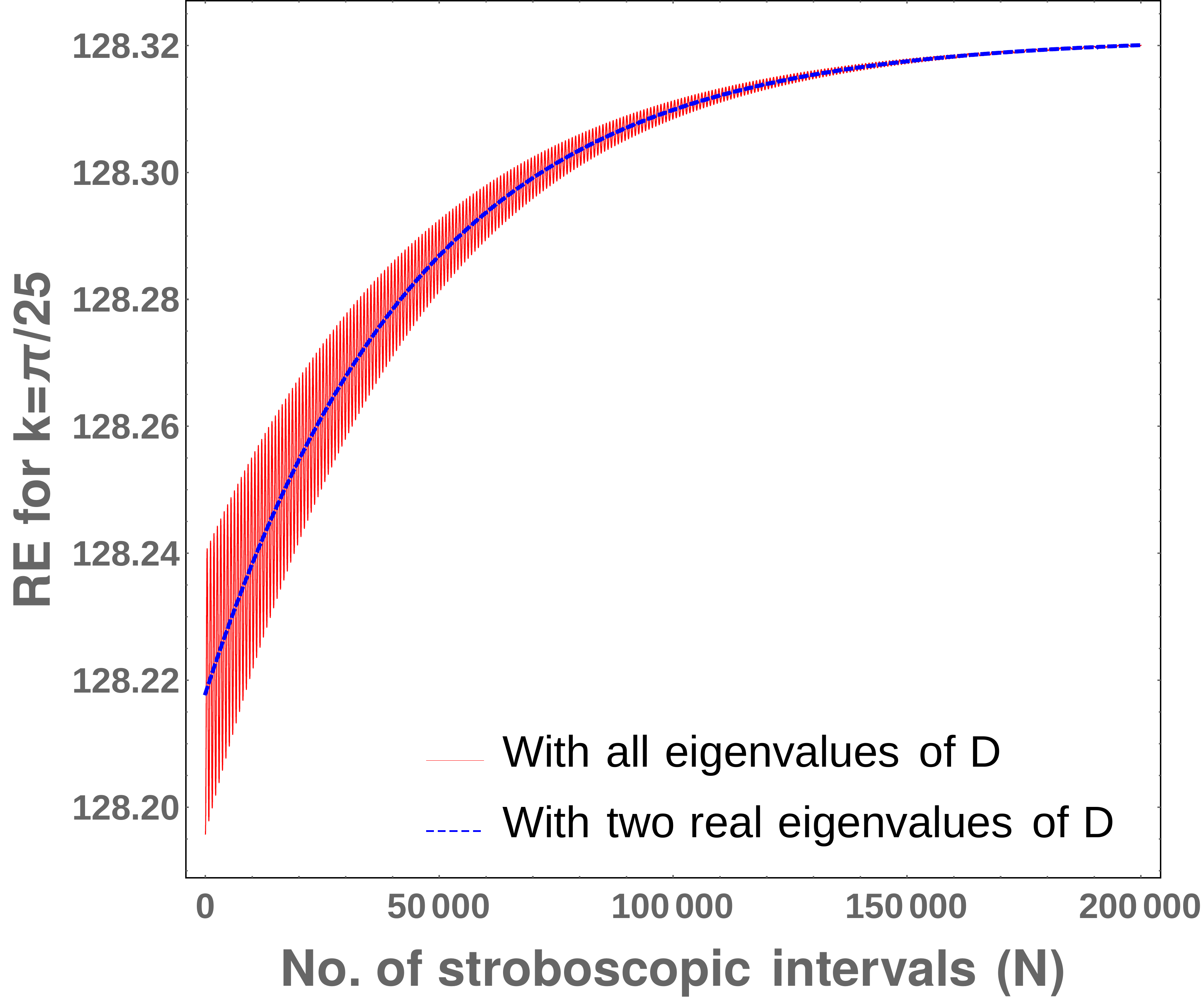}
		\label{fig_k5}}
	\hfill
	\quad
	\subfigure[]{%
		\includegraphics[width=.45\textwidth,height=6cm]{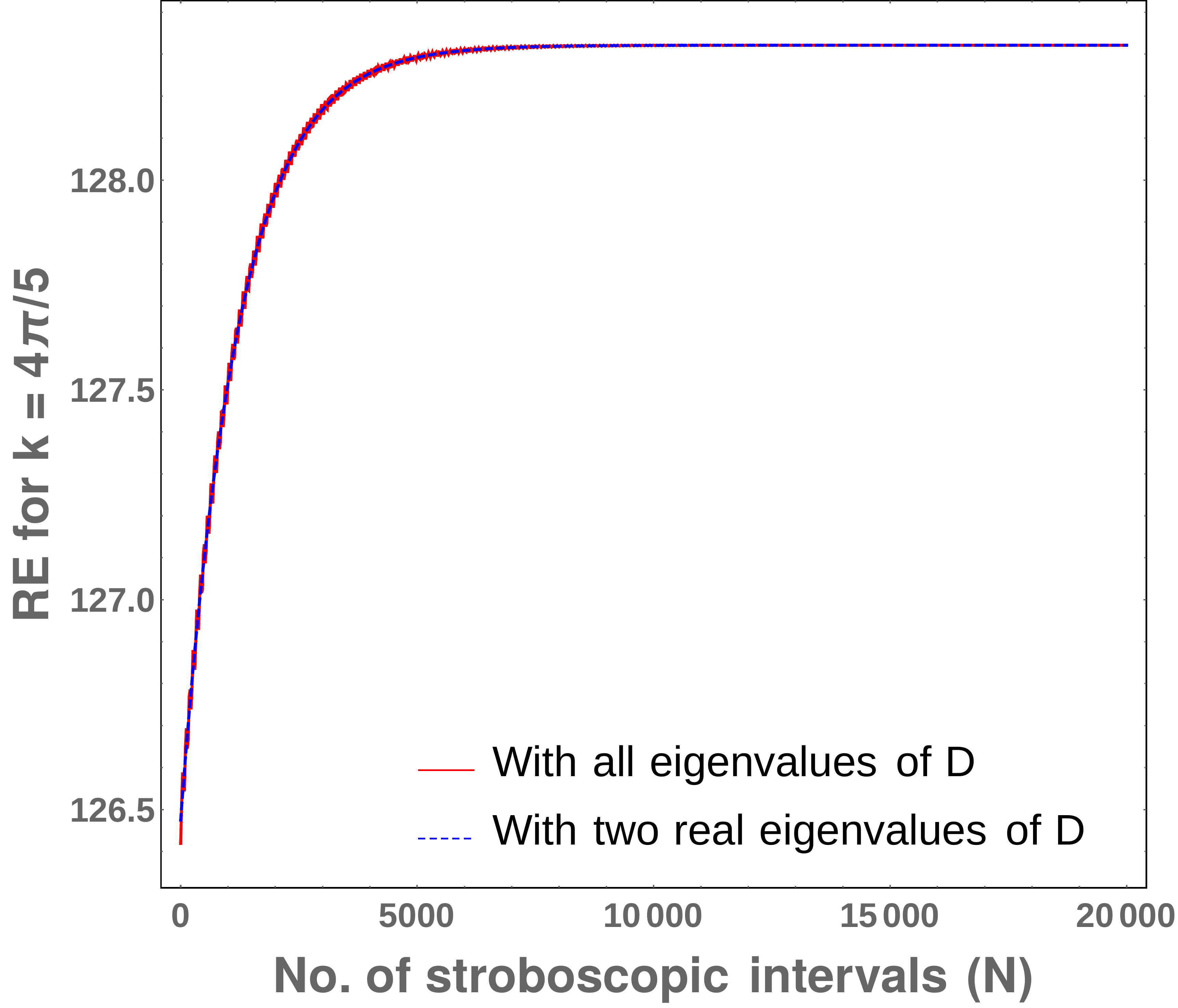}
		\label{fig_k75}}
	\caption{ (Color online) {The variation of the residual energy $\langle \varepsilon_{res}(k,NT)\rangle$ as a function of the stroboscopic periods $N$ for arbitary momentum modes $k=\pi/25$ and $k=4\pi/5$. (a) We see that the complex eigenvalues lead to rapid oscillations around the mean value determined by the real eigenvalues of the $D$-matrix for low $k=\pi/25$. These rapid oscillations as can be seen decay with increasing $N$ for any mode $k$. (b) These rapid oscillations are very small for an arbitrary high value of $k=4\pi/5$ justifying our assumptions in the main text.}}
	\label{fig_rapid}
\end{figure*}

At large N, the contribution of $[\lambda_c^\pm]^N=|\lambda_c(k,\alpha,\omega)|^N e^{\pm i N\phi(k,\alpha,\omega)}$ to the residual energy summed over all the $k$-modes vanish, as the interferences due to the fast oscillating phases $e^{\pm i N\phi(k,\alpha,\omega)}$ cancel each other. This is apparent from Fig. \ref{fig_rapid}, where for both the values of $k$, we see how the contributions of $[\lambda_c^\pm]^N$ to the exact RE$(k)$ (in red) oscillate around an increasing mean (in blue) set by contributions only from $1$ and $r^N$. For low $k$ (in Fig.~S\ref{fig_k5}) we observe that the rapid oscillations due to $[\lambda_c^\pm]^N$ in the RE$(k)$ sit on top of the mean, whereas, for high values of $k$ (see Fig.  S\ref{fig_k75}) the contributions of $[\lambda_c^\pm]^N$ to the RE$(k)$ are nearly zero and the mean approximately coincides with the exact RE$(k)$. Therefore, we can easily set the contributions of $[\lambda_c^\pm]^N$ as zero for each mode k throughout the rest of our calculations. However, $1$ and $r^N$ survives in such a large $N$ limit to yield,

\begin{equation}
D^N(k\neq 0,\pi)=SD_dS^{-1}=\frac{1}{2}\left(
\begin{array}{cccc}
1 & 0 & 0 & 1 \\
0 & 0 & 0 & 0 \\
0 & 0 & 0 & 0 \\
1 & 0 & 0 & 1 \\
\label{eq_dln}
\end{array}
\right)+\left[r(k,\alpha,\omega)\right]^N D_{NU}(k,\alpha,\omega)
\end{equation}
where $D_d=\operatorname{diag}\left(1,0,0,[r(k,\alpha,\omega)]^N\right)$.\\
Here the first matrix in the R.H.S is a constant matrix independent of the driving frequency, amplitude or protocol chosen and hence, universal. On the other hand, the second term in R.H.S. contains a non-universal matrix $D_{NU}(k,\alpha,\omega)$ which is of the form:
\begin{equation}
D_{NU}(k,\alpha,\omega)=\left(
\begin{array}{cccc}
u x_1(r) & v x_1(r) & v^* x_1(r) & -u x_1(r) \\
u x_2^*(r) & v x_2^*(r) & v^* x_2^*(r) & -u x_2^*(r) \\
u x_2(r) & v x_2(r) & v^* x_2(r) & -u x_2(r) \\
-u x_1(r) & -v x_1(r) & -v^* x_1(r) & u x_1(r) \\
\end{array}
\right)
\end{equation} 
where $u=u(k,\alpha,\omega)=\left[S^{-1}\right]_{41}$ and $v=v(k,\alpha,\omega)=\left[S^{-1}\right]_{42}$.\\

Although $D_{NU}(k,\alpha,\omega)$ specifically depends on the driving amplitude, frequency and the protocol implemented, it is essentially independent of the number of stroboscopic periods $N$ in this limit. All the time-dependence of the problem lies in the coefficient $\left[r(k,\alpha,\omega)\right]^N$ of the matrix $D_{NU}(k,\alpha,\omega)$. As this coefficient is positive and less than one, with increasing $N$, it gradually goes to zero only at $N\to\infty$ when the system finally attains its universal asymptotic value. But for any large non-zero $N<\infty$, this coefficient competes with the universal part (the first term in the R.H.S.) and generates a turning towards the asymptotic value (see Fig. 2 in the main text). This competition between the universal and the non-universal part is present irrespective of the protocol chosen, and hence, the curves for the residual energy in Fig. 2, are alike. Even though the slopes of the residual energy curves, depend upon both $r^N(k,\alpha,\omega)$ and $D^N(k,\alpha,\omega)$, the rate at which the curves for any protocol approach the asymptotic value with increasing $N$ is of course governed by the value of $r(k,\alpha,\omega)$  which is non-universal and varies depending upon the applied protocol.\\

\section{linear rise in the residual energy with low $N$} 
In this subsection,  we shall use approximate analytical methods  to show that for  small $p$ (and small $1-p$, where the case with $p=1$  corresponds to the perfectly periodic situation) and low $N$, the residual energy (RE)  (and also the difference in residual energy  from the periodic steady state value) increases  linearly with the number of stroboscopic periods $N$. We  shall also investigate the  variation of the RE  with $p$ for a given number of stroboscopic periods ($N$).   For the sake of  convenience in explaining the results, we shall restrict our attention only to the aperiodically $\delta$-kicked situation. Our argument below is  based on the notion of more probable configurations, for example if $p \to 0$, we shall consider the most probable configuration with  either no kick or the second most probable configurations with only one kick being  present in the entire process of driving.
Similarly, for  $(1-p) \to 0$,  there are only two more probable configurations that one can probe, namely, the configurations with 
only one kick  missing  and  the perfectly periodic configuration with $p=1$.

\begin{figure*}[b]
	\centering
	\subfigure[]{%
		\includegraphics[width=.45\textwidth,height=6cm]{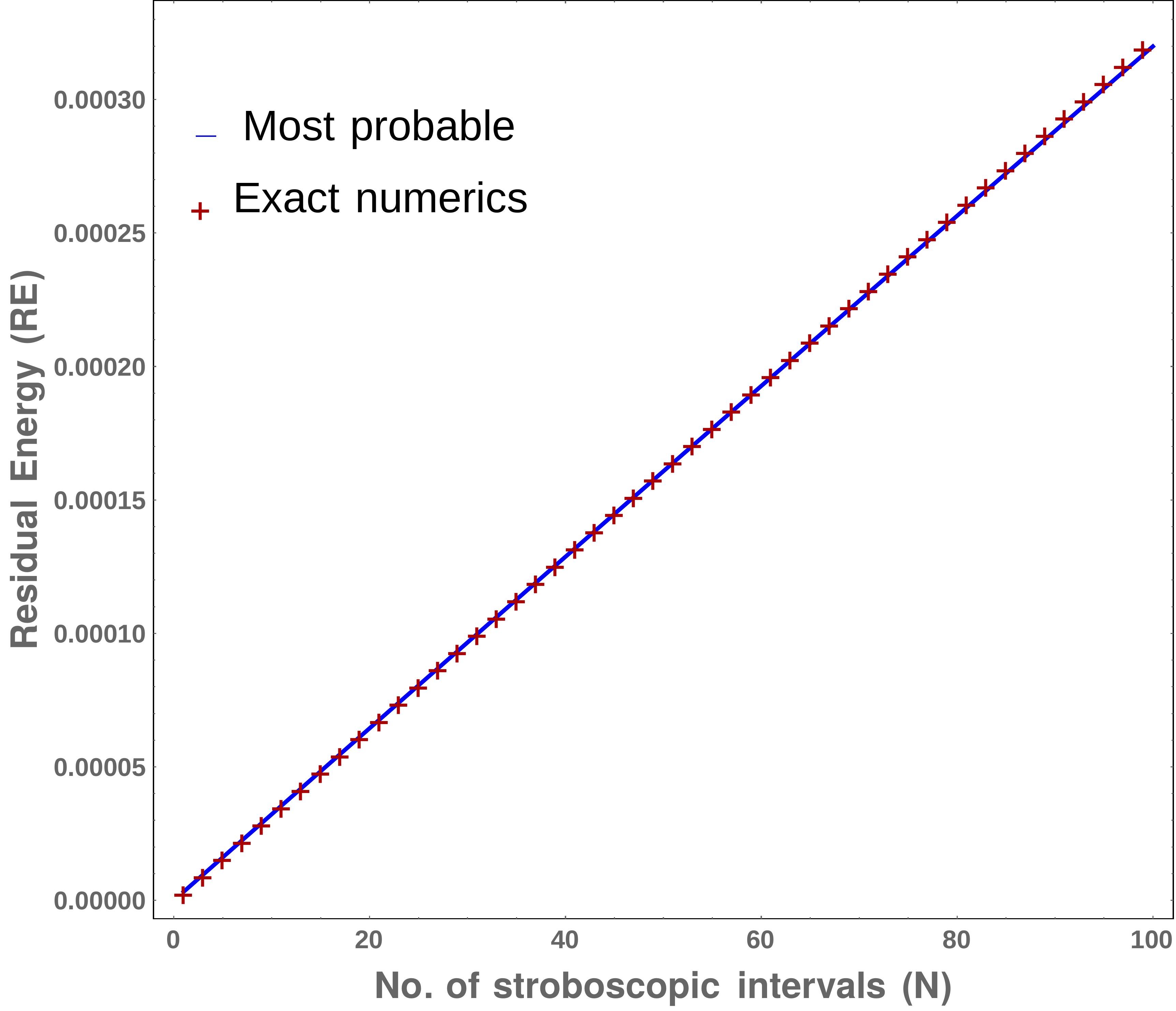}
		\label{fig_pt0}}
	\hfill
	\quad
	\subfigure[]{%
		\includegraphics[width=.45\textwidth,height=6cm]{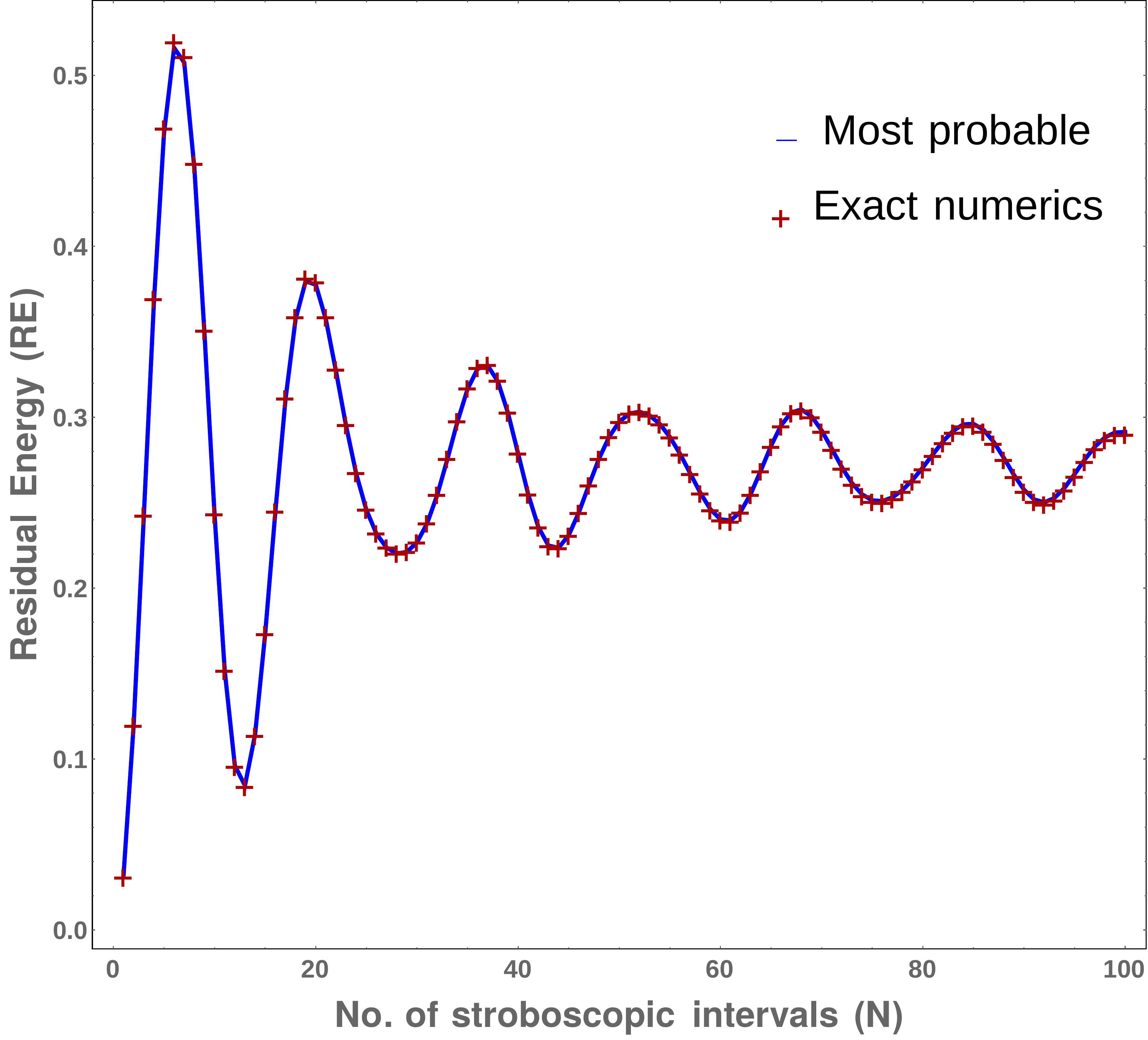}
		\label{fig_pt1}}
	\subfigure[]{%
		\includegraphics[width=.45\textwidth,height=6cm]{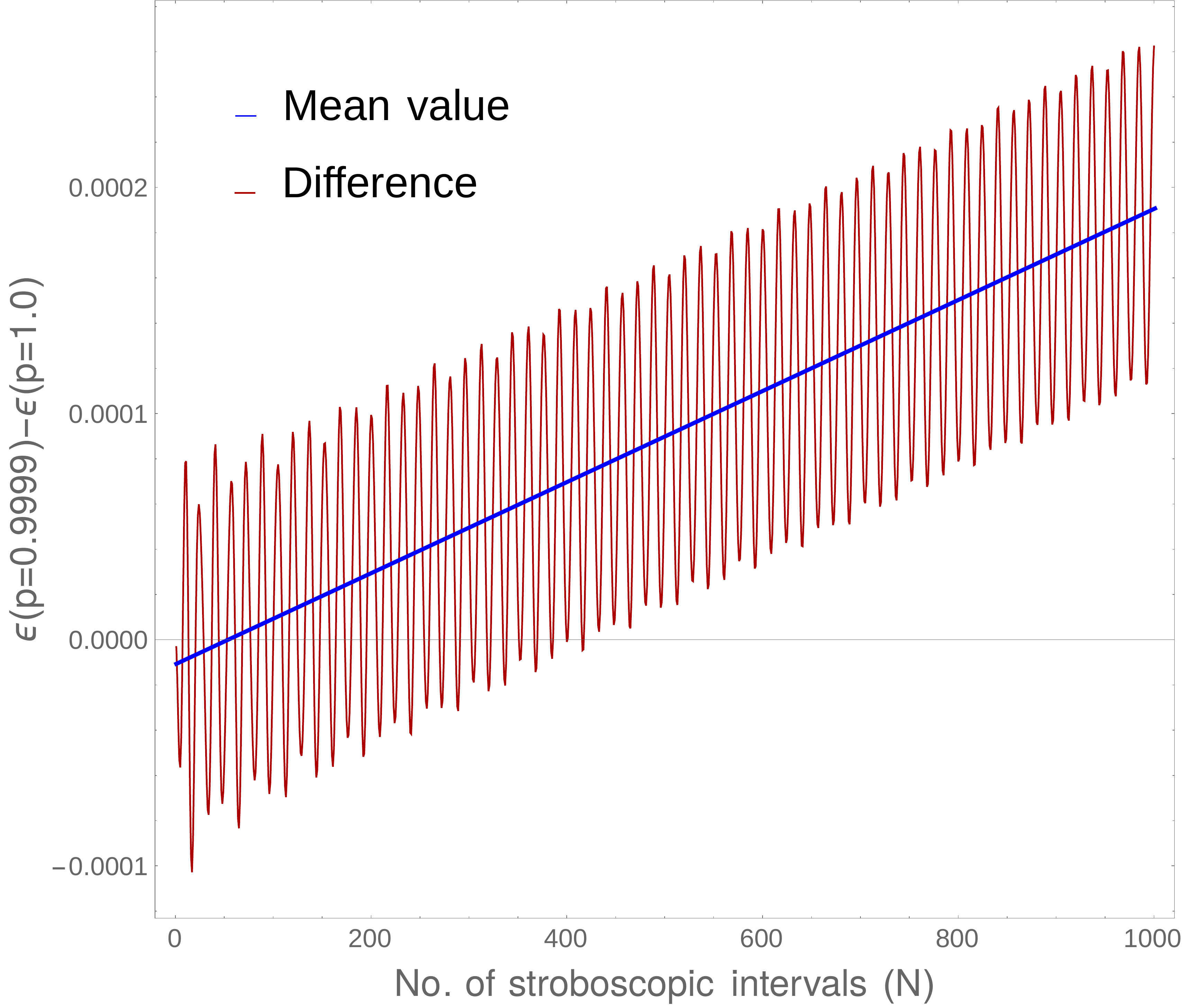}
		\label{fig_pt2}}
	\caption{ (Color online) {(a) The residual energy (RE) as obtained from the Eq.~\eqref{rese} (blue line) and exact numerical calculations (red cross) plotted as a function of $N$; we find that in the limit $p \to 0$, the RE grows linearly with $N$. We have chosen $p=0.0001$ and $\alpha=\pi/16$. (b) The RE as obtained from the Eq.~\eqref{eqpt1} summing over all possible one missed kick configurations are shown
			via the blue line, which is a perfect match with the exact numerical results plotted as red crosses over the blue line. Here, $p=0.9999$ and $\alpha=\pi/16$. (c) The mean value of the difference of the RE for the one missed kick situation shown in (b) and the perfectly periodic situation grows linearly with $N$.
		}}
		\label{fig_resvn}
	\end{figure*}

	Let us first consider the case with $p \to 0$ and further assume that there is no kick up to the stroboscopic time $mT$ (where $ 1 \leqslant m \leqslant N-1 $).  A kick is present at time
	$(m+1)T$; thereafter,  the system evolves freely up to the time
	$NT$. The evolved state at the final time $NT$ is then given by
	\begin{equation}
	\lvert \psi_k(NT) \rangle = \left[  U_k^{0}(T) \right]^{m}
	\mathcal{F}_k(T) \left[ U_k^0(T) \right]^{(N-m-1)} \lvert \psi(0) \rangle
	\end{equation}
	where $\boldmath{\cal{F}}_k(T)$ is the usual Floquet operator  and
	$U^0_{k}(T)= \exp(-i H^0_{k} T)$ is the time evolution operator for the
	free Hamiltonian $H^0_{k}$. For the $\delta$-kick situation, as discussed in the main text, we have an exact form of
	the Floquet operator,  $ \bold{\cal{F}}_k(T) = \exp(-i
	\alpha \sigma_z)\exp(-i H^0_{k}T) $. Using the Baker-Campbell-Hausdorff
	formula, one can easily find the expectation value
	\begin{eqnarray}
	e_k(NT) &=& \langle \psi_k(NT) \rvert H_k^0 \lvert \psi_k(NT) \rangle
	\nonumber \\
	&=& \langle \psi_k(0) \lvert \exp(i \alpha \sigma_z) H_k^{0} \exp(-i
	\alpha \sigma_z) \rvert \psi_k(0) \rangle \equiv e_k
	\end{eqnarray}
	
	Let us recall that  the ground state of the Hamiltonian $H^0_k=(1-\cos k)\sigma_z + (\sin
	k )\sigma_x$ can be written as $\lvert \psi_k(0) \rangle = (-f_{-} , f_{+}
	)^{T}$ , where $f_{\pm} = \sqrt{1/2(1 \pm \sqrt{(1-\cos k)/2})}$ with the corresponding
	ground state energy $e^g_k(0)= - 2 \sin(k/2) $. Using the identity  $\exp(-i \alpha
	\sigma_z) = e^{-i \alpha} \lvert + \rangle \langle + \rvert +  e^{i
		\alpha} \lvert - \rangle \langle - \rvert$ , where $\lvert+\rangle =
	(1,0)^{T} $ and $ \lvert-\rangle = (0,1)^{T} $, and the relations $(f_{-}^2 -f_{+}^{2})=\sin (k/2)$ and $f_+f_- =
	(1/2)\cos(k/2)$,  we readily arrive at the expression
	\begin{eqnarray}
	e_k = (1-\cos k)(f_{-}^2 -f_{+}^{2}) -2\sin k \cos (2\alpha) (f_+f_-)
	= (\cos k -1) \sin(k/2) - \sin k \cos(k/2) \cos(2\alpha)
	\end{eqnarray}
	Finally, we find the expression for  the RE,
	\begin{eqnarray}
	\mathcal{\epsilon}_{res} &=& \int_{0}^{\pi} \frac{dk}{\pi} \left[
	e_k - e^g_k(0) \right]=  \int_{0}^{\pi} \frac{dk}{\pi}
	\left[ (\cos k -1) \sin(k/2) - \sin k \cos(k/2) \cos(2\alpha) + 2\sin
	(k/2)
	\right] \nonumber =  \frac{8}{3\pi} \sin^2 \alpha.
	\label{eres}
	\end{eqnarray}

	What is important is that in the situation when only one kick is present in the entire driving (i.e., $p \to 0$), the RE is independent of the stroboscopic instant $m$ at which the kick is applied  and further it is entirely  determined by  the strength of the kick. In this situation,  the configuration averaged RE  can be written as,
	
	\begin{eqnarray}
	\varepsilon_{res}(NT) = \langle \epsilon_{res} \rangle = \binom{N}{0}(1-p)^N \epsilon_{res}^{(0)} + \binom{N}{1}p(1-p)^{N-1} \epsilon_{res}^{(1)}
	\label{eqpt0}
	\end{eqnarray}
	where the first term corresponds to the no-kick situation whose contribution to the RE  is zero, whereas the second term provides a non-zero contribution due to the presence of a single kick. Let us note that  we have neglected the configurations with a higher number of kicks which occur with a vanishingly small probability in the limit $p \to 0$ and $N \to \infty$.  Finally, in the limit $p \rightarrow 0$, we get the RE  as
	\begin{eqnarray}
	\varepsilon_{res}(NT)= \frac{8Np(1-p)^{(N-1)}}{3\pi} \sin^2\alpha.
	\label{rese}
	\end{eqnarray} 
	Remarkably, the RE  grows  linearly  with $N$ as shown in Fig.~\ref{fig_pt0}.
	
	In the other limit,  $(1-p) \rightarrow 0$, considering the two more probable terms in the configuration averaged RE of the system, one can similarly  write,
	\begin{equation}
	\varepsilon_{res}(NT) = \binom{N}{N}(p)^N \epsilon_{res}^{(N)} + \binom{N}{N-1}p^{(N-1)}(1-p) \epsilon_{res}^{(N-1)}.
	\label{eqpt1}
	\end{equation}
	Notably, $\epsilon_{res}^{(N)}$ corresponds to the RE  of the system for the perfectly periodic driving (as derived in the main text)  and $\epsilon_{res}^{(N-1)}$ is the RE when only one kick is missing in the entire driving. The exercise to arrive at an analytic expression in this case is tedious and unilluminating, more because unlike the previous situation  the quantity $\epsilon_{res}^{(N-1)}$ involves the sum over the stroboscopic instants at which the  kick is missed.  To circumvent this problem,  we shall  average over the all possible permutations i.e. sum over  $m$ (position of  the missed kick, $1 < m < (N-1)$) with an equal weight $1/N$ (see Fig.~\ref{fig_pt1}). Henceforth, we see that the difference  in the RE for the one  missed kick situation and the perfectly periodic  situation 
	(i.e.,  the contribution to the RE that destabilizes the periodic steady state) indeed grows  linearly with $N$ as shown in Fig.~\ref{fig_pt2} upto an appropriate value of $N$ for which the approximation of single drive and one drive cycle missing holds.

	Finally, we address the question how does variation of the  RE (after a given number of stroboscopic periods) depend on the probability $p$; this however depends
	on the driving frequency $\omega$ and amplitude $\alpha$. In Fig. \ref{fig_revspfr} and Fig.~\ref{fig_revspal2}, we show that the variation is neither monotonic nor  symmetric with $p$.
	
	\begin{figure*}[h]
		\centering
		\subfigure[]{%
			\includegraphics[width=.45\textwidth,height=6cm]{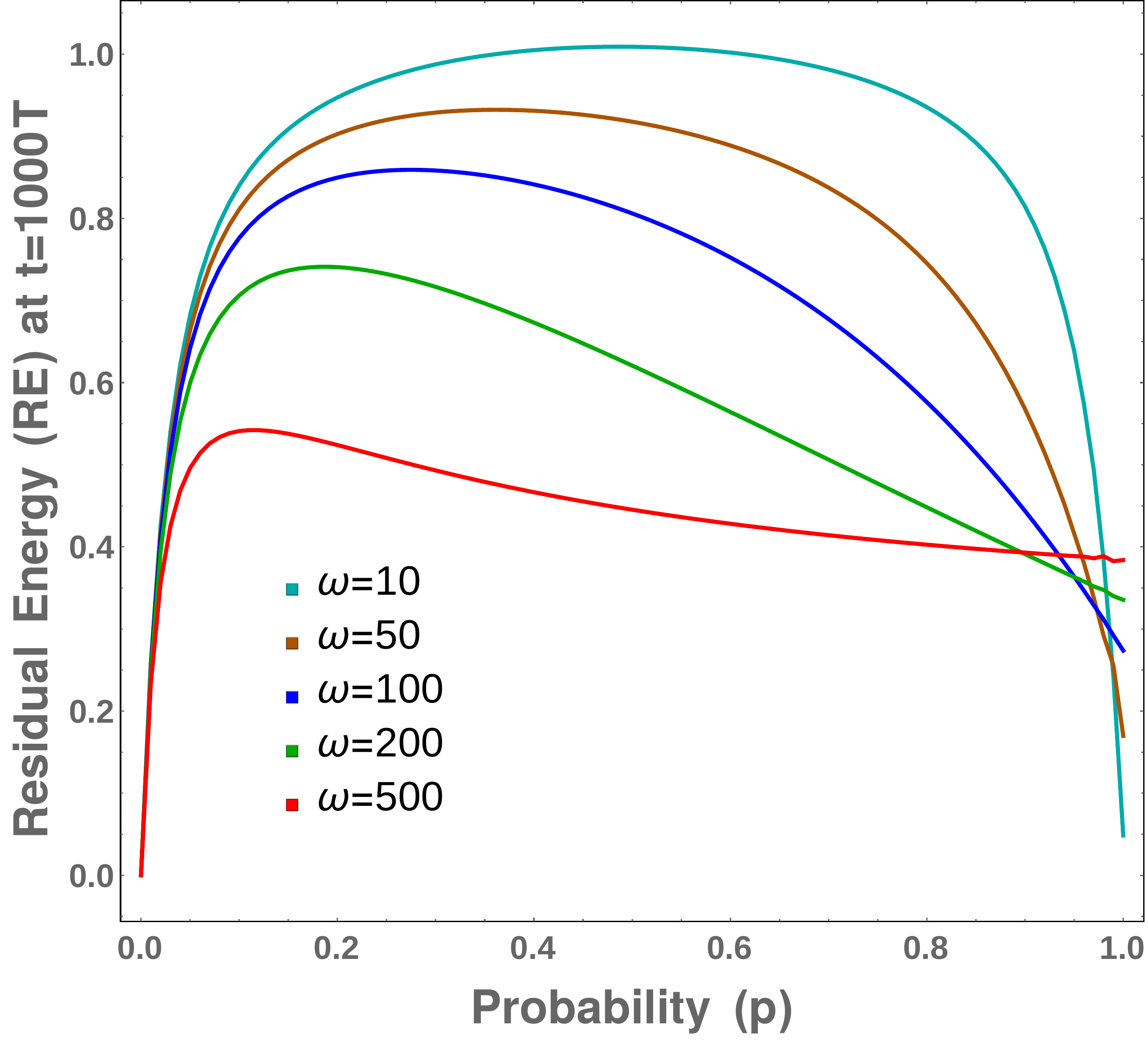}
			\label{fig_revspfr}}
		\hfill
		\quad
		\subfigure[]{%
			\includegraphics[width=.45\textwidth,height=6cm]{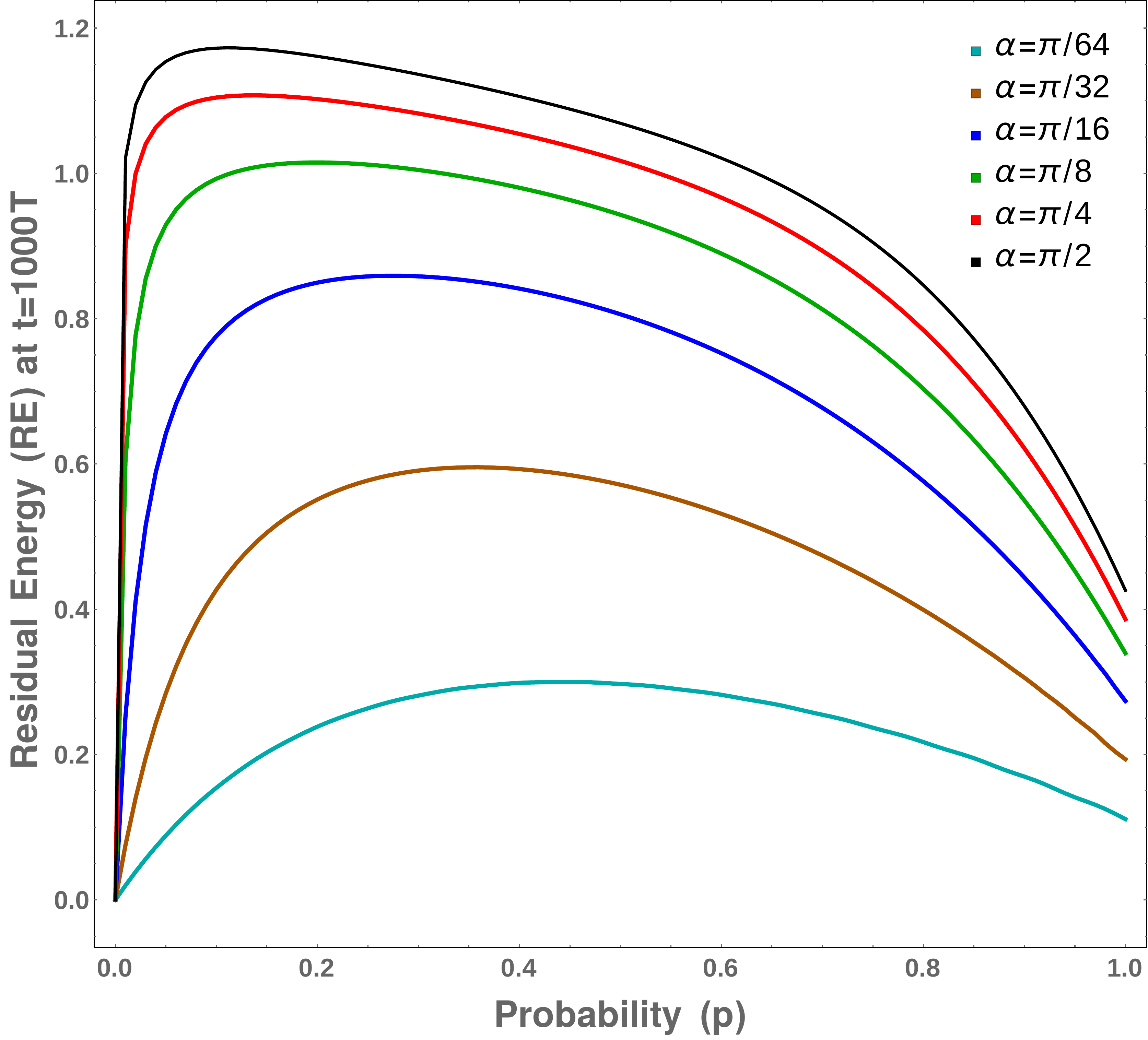}
			\label{fig_revspal2}}
		\caption{ (Color online) { The variation of the RE as obtained using exact numerical methods after a given number of stroboscopic periods ($N=1000$) as function of $p$: (a) for different values of  $\omega$ with $\alpha = \pi /16$ and (b) for different values of $ \alpha$ with $\omega=100$. In both the situations, we find that the variation is neither monotonic nor  symmetric with $p$.
			}}
			
		\end{figure*}

\end{document}